\documentclass[aps,prl,twocolumn,showpacs,floatfix]{revtex4-1}

\usepackage{bm,dcolumn,amsmath,graphicx}
\usepackage{epsfig}
\usepackage{nicefrac}
\usepackage{multirow}
\usepackage{longtable}

\newcommand{\Fig}[1]{Fig.~\ref{#1}}

\newcommand{\E}[1]{\ensuremath{\times 10^{#1}}}
\newcommand{\cm}{\ensuremath{\textrm{cm}^{-1}}}

\newcommand{\FeV}{\textrm{Fe\,V}}
\newcommand{\NiV}{\textrm{Ni\,V}}

\newcommand{\Qa}{\ensuremath{Q_\alpha}} 
\newcommand{\ka}{\ensuremath{k_\alpha}} 
\newcommand{\daa}{\ensuremath{\Delta\alpha/\alpha}}
\newcommand{\val}{\ensuremath{n}}

\begin{document}

\title{Limits on the dependence of the fine-structure constant on
gravitational potential from white-dwarf spectra}

\author{J.~C.~Berengut}
\author{V.~V.~Flambaum}
\author{A.~Ong}
\author{J.~K.~Webb}
\affiliation{School of Physics, University of New South Wales, Sydney, NSW 2052, Australia}

\author{John~D.~Barrow}
\affiliation{DAMTP, Centre for Mathematical Sciences, University of Cambridge, Cambridge CB3 0WA, United Kingdom}

\author{M.~A.~Barstow}
\author{S.~P.~Preval}
\affiliation{Department of Physics and Astronomy, University of Leicester, University Road, Leicester LE1 7RH, United Kingdom}

\author{J.~B.~Holberg}
\affiliation{Lunar and Planetary Laboratory, Sonett Space Science Building, University of Arizona, Tucson, Arizona 85721, USA}

\date{9 July 2013}

\begin{abstract}

We propose a new probe of the dependence of the fine structure constant, $\alpha$, on a strong gravitational field using metal lines in the spectra of white dwarf stars. Comparison of laboratory spectra with far-UV astronomical spectra from the white dwarf star G191-B2B recorded by the Hubble Space Telescope Imaging Spectrograph gives limits of $\Delta \alpha /\alpha =(4.2\pm 1.6)\times 10^{-5}$ and $(-6.1\pm 5.8)\times 10^{-5}$ from \FeV\ and \NiV\ spectra, respectively, at a dimensionless gravitational potential relative to Earth of $\Delta \phi \approx 5\times 10^{-5}$. With better determinations of the laboratory wavelengths of the lines employed these results could be improved by up to two orders of magnitude.

\end{abstract}

\maketitle

Light scalar fields can appear very naturally in modern cosmological models
and theories of high-energy physics, changing parameters of the Standard
Model such as fundamental coupling constants and mass ratios. Like the
gravitational charge, the scalar charge is purely additive, so near massive
objects such as white dwarfs the effect of the scalar field can change. For
objects that are not too relativistic, like stars and planets, both the
total mass and the total scalar charge are simply proportional to the number
of nucleons in the object. However, different types of coupling between the
scalar field and other fields can lead to an increase or decrease in scalar
coupling strengths near gravitating massive bodies~\cite{mbs}. For small
variations, the scalar field variation at distance $r$ from such an object
of mass $M$ is proportional to the change in dimensionless
gravitational potential $\phi=GM/rc^2$,
and we express this proportionality by introducing the sensitivity
parameter, $k_{\alpha }$~\cite{flambaum08aipconf}. Specifically, for
changes in the fine structure `constant', $\alpha $, we write
\begin{equation*}
\Delta \alpha /\alpha \equiv \frac{\alpha (r)-\alpha _{0}}{\alpha _{0}}%
\equiv k_{\alpha }\,\Delta \phi =k_{\alpha }\,\Delta \left( \frac{GM}{rc^2}%
\right) \,.
\end{equation*}

This dependence can be seen explicitly in particular theories of varying $%
\alpha $, like those of Bekenstein \cite{bek} and Barrow-Sandvik-Magueijo 
\cite{bsbm}, and their generalisations \cite{blip}, where $\alpha $ can
increase ($\Delta \alpha /\alpha >0$) or decrease ($\Delta \alpha /\alpha <0$%
) on approach to a massive object depending on the balance between
electrostatic and magnetic energy in the ambient matter fields \cite{mbs}.
The most sensitive current limits on $k_{\alpha }$\ come from measurements
of two Earth-bound clocks over the course of a year~\cite%
{flambaum08aipconf,bauch02prd,ferrell07pra,fortier07prl,blatt08prl,guena12prl,BShaw,leefer13arxiv}%
. The sensitivity is entirely due to ellipticity in the Earth's orbit, 
which gives a 3\%\ seasonal variation in the gravitational potential
at the Earth due to the Sun. The peak-to-trough 
sinusoidal change in the potential has magnitude $\Delta\phi =3\times 10^{-10}$.
Each clock has a different sensitivity to $\alpha $%
-variation, and so $\Delta \alpha /\alpha $\ can be measured and hence $%
k_{\alpha }$\ extracted.

Due to the high precision of atomic clocks, $k_{\alpha }$ is determined very
precisely despite the relatively small seasonal change in the gravitational
potential. By contrast, we examine a `medium strength' field, where $\Delta \phi $
is five orders of magnitude larger than in the Earth-bound experiments, and
the distance between the probe and the source is $\sim 10^{4}$ times smaller
than 1\,AU. This allows us to probe nonlinear coupling of $\Delta \alpha
/\alpha $\ on $\Delta \phi $, or the effects of a scalar charge $Q$ which
produces a Yukawa-like scalar field $\Phi =Qe^{-mr}/r$ where $m$ is the
(very small) mass of the scalar. 

In this work we use the high-resolution far-UV spectrum of the nearby
($\approx 45$ pc~\cite{reid88apj}), hot
hydrogen-rich (DA) white dwarf G191-B2B, recorded by the Hubble Space
Telescope Imaging Spectrograph (STIS), which contains several hundred
absorption lines identified as \FeV\ and \NiV\ 
transitions~\cite{preval13mnras}. These iron and nickel
ions reside in the atmosphere of the white dwarf and the observed features
are formed in its outer layers, near the surface of white dwarf.
Consequently, the ions experience the strong downward surface gravity
of the star, $\log g=7.53\pm 0.09$, but are
supported against this by the transfer of momentum from high-energy photons,
a process termed ``radiative levitation''~\cite{chayer95apj}.
Here $g = GM/R^2$ in cgs units, with
$M_\textrm{WD} = 0.51\,M_\odot$ and 
$R_\textrm{WD} = 0.022\,R_\odot$~\cite{preval13mnras}.
The gravitational potential for ions in the atmosphere
of this white dwarf relative to the laboratory is 
$\Delta \phi \approx 5\times 10^{-5}$.

\begin{figure*}[tb]
\includegraphics[width=0.8\textwidth]{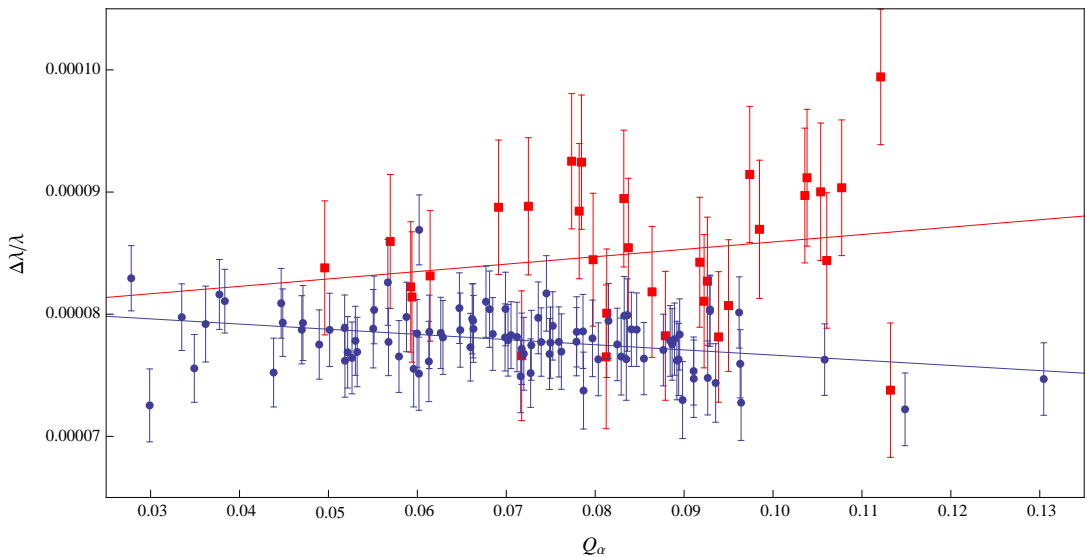}
\caption{$\Delta\protect\lambda/\protect\lambda$\ vs. $Q_\protect\alpha$\
for transitions in \FeV\ (blue circles) and \NiV\ (red
squares). The slope of the lines give $\Delta\protect\alpha/\protect\alpha  %
= (4.2\pm1.6) \times 10^{-5} $ and $(-6.1\pm5.8)\times 10^{-5} $ for \textrm{%
Fe\,V}\ and \NiV, respectively. The slope seen in the \NiV\ spectra
is likely due to systematics present in the laboratory wavelength measurements, 
rather than indicating a gravitational dependence of $\alpha$.
}
\label{fig:FeNi}
\end{figure*}

To extract dependence on any $\alpha$ variation we first calculate the
sensitivity coefficient for each line. As in previous work, we parameterize
the sensitivity of the transition frequency to a variation in $\alpha $ from
the laboratory value $\alpha _{0}$ by the $q$-coefficient, defined in terms
of the line frequency $\omega $ by 
\begin{equation}
q=\left. \frac{d\omega }{dx}\right\vert _{x=0}\,,  \label{eq:q_def}
\end{equation}%
where $x\equiv (\alpha /\alpha _{0})^{2}-1\approx 2\Delta \alpha /\alpha $
is the fractional (small) change in $\alpha ^{2}$. The frequencies of lines
that are observed in the white dwarf spectra are shifted from their
laboratory values, $\omega _{0}$, due to the sum of Doppler and
gravitational redshifts, $z$, and any potential gravitational $\alpha $ dependence
near the white dwarf: 
\begin{equation*}
1+z=\frac{\omega _{0}+qx}{\omega }\,.
\end{equation*}%
The relationship between the laboratory wavelengths and those observed near
the white dwarf is 
\begin{equation}
\frac{\Delta \lambda }{\lambda _{0}}=\frac{\lambda -\lambda _{0}}{\lambda
_{0}}=z-Q_{\alpha }\frac{\Delta \alpha }{\alpha }(1+z)\,,  \label{lambda}
\end{equation}%
where $Q_{\alpha }=2q/\omega _{0}$ is the relative sensitivity of the
transition frequency to variation in $\alpha $. In Fig.~\ref{fig:FeNi} we
present a graph of $\Delta \lambda /\lambda $\ vs. $Q_{\alpha }$\ for both 
\FeV\ lines (blue circles) and \NiV\ lines (red squares). The data
used to generate these graphs can be found in Tables~\ref{tab:linesFe} and
\ref{tab:linesNi}.

To determine the sensitivity coefficients, $q$, for each line, we perform an 
\emph{ab initio} calculation of the spectrum for $x=-0.01$, $0.0$, and $0.01$%
, and then extract $q$ using~(\ref{eq:q_def}). The spectrum is calculated
using the CI+MBPT method~\cite{dzuba96pra}, a combination of
configuration-interaction and many-body perturbation theory. Details of the
implementation can be found in~\cite%
{berengut05pra,berengut06pra,berengut08jpb}. Here, we outline only the
important points and defer details to a later work. The final $q$ values are
presented in Tables~\ref{tab:linesFe} and \ref{tab:linesNi} for \FeV\ and \NiV,
respectively.

For both \FeV\ and \NiV\ we start with
a Dirac-Fock calculation in the $V^{N}$ potential (i.e. the self-consistent
field of all electrons) including the valence configuration $3d^{\val}$ where $%
\val=4$ for \FeV\ and $\val=6$ for \NiV. In
this procedure we simply scale the Dirac-Fock potential of the filled $%
3d^{10}$ shell by the number of valence electrons. We then form a $B$-spline
basis by diagonalizing a set of splines over the self-consistent potential,
which we use to form configurations with specified total angular momenta
for our configuration interaction (CI) calculation.
Configurations are formed by taking single and double excitations from the 
leading configurations $3d^{\val}$, $3d^{\val-1}\,4s$ and $3d^{\val-1}\,4p$. 
In the case of \FeV\ we use a $B$-spline basis of size $11spdf7g$ and
include all single and double excitations from the leading configurations.
The resulting energy levels are sufficiently close (within $\sim2\%$) to the
available data~\cite{sugar85jpcrd}.

For \NiV, the number of valence orbitals used for the CI
calculation is markedly smaller. We include single-electron excitations to $%
12spdf$ and double excitations up to $5spdf$ from the leading configurations
(a similar strategy was used for Cr\,II in~\cite{berengut11pra1}). Results
using all single and double excitations to $7s6pdf$ were consistent,
although the final energies were not as good. Many-body perturbation theory
(MBPT) corrections using a valence basis of $30spdfgh$ were then added to
the CI calculations, which improve the overall agreement with the experimental
values (again level energies are within $\sim2\%$).
Note that while the $q$ values themselves do not change very much with the
addition of MBPT corrections for either ion, the energy levels are much better
and this helps with their identification.

The Hubble Space Telescope (HST) STIS spectrum utilized in this work is unique in coupling the
highest signal-to-noise so far achieved for any white dwarf with the best
spectral resolution and spectral coverage available with the instrument. The
spectrum is constructed from a series of high-resolution 
(resolving power $R\approx 144,000$) 
observations obtained with the E140H and E230H gratings as part of an
extensive calibration programme for the instrument, designed to provide flux
calibration at the 1\% level for all E140H and E230H primary and secondary
echelle grating modes~\cite{holberg03aspcs}. The detailed list of
observations and their reduction, merging and co-adding the components, has
been reported by \cite{preval13mnras}.

In summary, the outcome of this work
was two single continuous spectra spanning the wavelength ranges 1160--1680%
\AA\ and 1625--3145\AA\ for E140H and E230H respectively. The signal-to-noise
is typically $\approx 50$ but exceeds $100$ at some wavelengths.

These spectra contain almost 1000 absorption features, mostly in the
1160--1680\AA\ region. Cross-correlating their measured wavelengths with
lines from the Kurucz~\cite{kurucz11cjp} and Kentucky~\cite{kentuckylinelist}
lists yields 914 identifications. A large number of these correspond to 
\FeV\ and \NiV\ transitions. The detailed identification
work has been reported by~\cite{preval13mnras}.

\FeV: Of the original 106 \FeV\
transitions identified in the HST spectra, there are 96 for which there are
good laboratory wavelengths (taken from~\cite{ekberg75pscr}).
Ref.~\cite{ekberg75pscr} estimate an
uncertainty of $0.004$\,\AA\ in their measurements, which dominates
the errors for each value of $\Delta \lambda /\lambda $. From Fig.~\ref%
{fig:FeNi} we extract $\Delta \alpha /\alpha =(4.2\pm 1.6)\times 10^{-5}$
(an apparent $2.6\sigma$ deviation from zero).

It is interesting to note that statistically the laboratory errors seem to
be overestimated. Simply comparing the laboratory wavelengths with the HST
data suggests that the actual error in the laboratory data is $\sim3$\,m\AA,
rather than the claimed 4\,m\AA. 
An unweighted fit from Fig.~\ref{fig:FeNi} gives $\Delta\alpha/\alpha  =
(4.3\pm1.2)\times 10^{-5}$.

One potential source of systematic error is the calibration of the
laboratory measurements or the astronomical data. Offset errors will not
affect our result, since this simply causes a change in the measured doppler
shift, $z$. On the other hand gain calibration errors --- a linear mapping
between the real and the measured $\lambda $ --- \emph{could} cause a
spurious detection of gravitational $\alpha$-dependence \emph{if} 
there is also a correlation between $Q_{\alpha}$\ and $\lambda$.
If there is no correlation, then any gain
error would not matter since the data points would be completely randomised
on the $Q_{\alpha }$-axis.

In fact, such a correlation does exist. Energy levels of an ion that have a
larger binding energy tend to spend more time closer to the nucleus, and
therefore have larger relativistic effects. Therefore higher energy
transitions (smaller $\lambda$) will tend to have a larger difference in the
relativistic effects between the upper and lower levels, and hence have
larger $q$. The correlation between $q$ and $\lambda_0$ is $-0.45$ for the 
\FeV\ lines used (there is no evidence of non-linear
correlations). Therefore, gain shift in the laboratory measurements or the
calibration of the HST spectrograph would be a possible source of error.

\begin{figure*}[tb]
\includegraphics[width=0.8\textwidth]{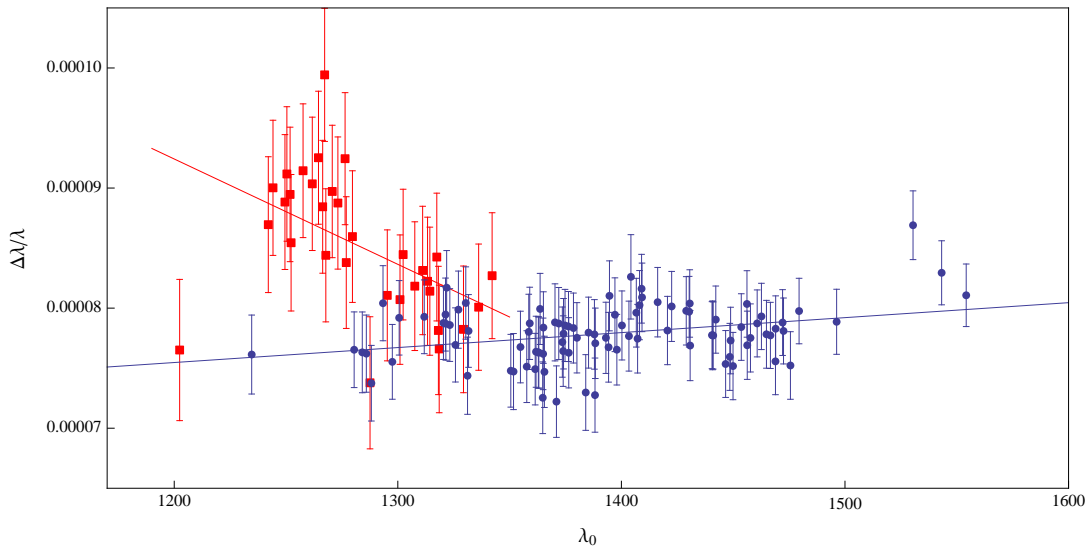}
\caption{$\Delta\protect\lambda/\protect\lambda$\ vs. $\protect\lambda_0$
for transitions in \FeV\ (blue circles) and \NiV\ (red squares).
The correlation seen here could be due to calibration systematics, or
gravitational dependence of $\alpha$ since \Qa\ and $\lambda_0$ are (anti-)correlated.
}
\label{fig:FeNi_wavelength}
\end{figure*}

We can account for the potentially spurious detection of $\alpha $-variation that may
occur by first removing any linear dependence of $\Delta \lambda /\lambda $\
on $\lambda $ (see Fig.~\ref{fig:FeNi_wavelength}). The line of best fit for 
\FeV\ (blue) in Fig.~\ref{fig:FeNi_wavelength} is $%
(\Delta \lambda /\lambda )_{\textsl{model}}=7.79\times 10^{-5}+1.25\times
10^{-8}\,(\lambda _{0}-1394\,\text{\AA })$. 
Here, the first term (see eq.(\ref{lambda})), $z_{abs}=7.79\times 10^{-5},$
is the average total redshift of the \FeV\ lines. The
fitted model values, $(\Delta \lambda /\lambda )_{\textsl{model}}$ as a
function of $\lambda _{0}$, are removed from the observed values of $\Delta
\lambda /\lambda $\ and we plot these against $Q_{\alpha }$\ to obtain a new
value of $\Delta \alpha /\alpha =(2.8\pm 1.6)\times 10^{-5}$, consistent
with zero at the $1.77\,\sigma $ level (although again we note that if we
reduce the assumed laboratory errors to the level suggested by the data, the
error in $\Delta \alpha /\alpha $\ is of order $1.2\times 10^{-5}$). Note
that while we have removed the potential systematic due to calibration
error, we have also potentially lost a real signal of $\Delta \alpha /\alpha 
$. Ultimately, well calibrated laboratory and astronomical data will remove
the need for this procedure and boost the sensitivity of this method.

\NiV: Laboratory data for \NiV\ is
provided by~\cite{raassen77physc}, who estimate their uncertainty as 
$\sim 1$\,m\AA~\cite{raassen76physc}. In fact, based on comparison with the HST data,
this seems likely to be an underestimate. In Fig.~\ref{fig:FeNi} we use an
assumed laboratory error of 7\,m\AA , which leads to a more realistic
distribution of residuals. Calculations for \NiV\ are
also more difficult than for \FeV, and in several cases
potentially useful transitions were not used because the levels could not be
uniquely identified in our calculations. In other cases the original
lab data were blended (which, aside from being flagged in \cite%
{raassen77physc}, lead to obvious $>3\sigma $ outliers in Fig.~\ref{fig:FeNi}%
). In total, 32 \NiV\ transitions were used out of the 44 identified in HST spectra.
The slope of the line in \Fig{fig:FeNi} gives a value of 
$\Delta \alpha /\alpha =(-6.1\pm 5.8)\times 10^{-5}$, 
consistent with zero at the $1.05\,\sigma $ level.

As we did for \FeV, we removed any potential gain-shift
systematic by subtracting the linear dependence of $\Delta \lambda /\lambda $%
\ on $\lambda $: $(\Delta \lambda /\lambda )_{\textsl{model}}=8.51\times
10^{-5}-8.77\times 10^{-8}(\lambda _{0}-1283\,\text{\AA })$ 
(red line in Fig.~\ref{fig:FeNi_wavelength}). Note that for \NiV\ 
the slope is in the opposite direction to that of \FeV\ 
and is much larger (that is, $\Delta \lambda /\lambda $ 
tends to be smaller in transitions with longer wavelengths). Plotting $%
\Delta \lambda /\lambda -(\Delta \lambda /\lambda )_{\textsl{model}}$
against $Q_{\alpha }$\ gives a line of best fit 
$\Delta \alpha /\alpha =(-2.5\pm 5.8)\times 10^{-5}$, within $0.4\sigma$
of zero.

The two values for \daa\ that we have obtained (shown in \Fig{fig:FeNi}) 
seem inconsistent: the \FeV\ data indicates $\daa>0$, while the \NiV\ data 
indicates $\daa<0$. Formally, the weighted mean of the two values is 
$\daa = (3.5\pm 1.5)\E{-5}$, which is dominated by the \FeV\ result and
consistent with the \NiV\ result at $1.6\sigma$. 
\Fig{fig:FeNi_wavelength} suggests that the slope seen in the \NiV\ data 
is a systematic present in the laboratory wavelength measurements, rather than indicating 
a gravitational dependence of $\alpha$. Indeed, removing this potential systematic 
as described above brings both sets of data into agreement within $1\sigma$.

The linear dependence of $\alpha$ on gravitational potential derived from the 
\FeV\ measurement is $\ka = 0.7 \pm 0.3$, which is much weaker than the limit 
derived from atomic clocks, where the best current limit is 
$\ka = (-5.5\pm5.2)\E{-7}$~\cite{leefer13arxiv}. The aim of this work, however,
is not to find \ka, but to find the dependence of alpha on gravitational potential 
mediated by a light scalar field.
The white dwarf result probes a `medium field' limit, where the change in
dimensionless gravitational potential is five orders of magnitude larger that probed
using clocks, and the distance between the source and the probe is four orders of magnitude smaller.
The limit on \daa\ derived from analysis of white-dwarf spectra may be 
more sensitive to nonlinear coupling of scalar fields to $\alpha$, 
or $\alpha$-dependence due to a Yukawa-like scalar field with non-zero scalar mass.

We can compare the limits on the change in $\alpha$ measured in this system
with the application of the many-multiplet method to a quasar absorption
system~\cite{dzuba99prl,webb99prl}, which typically are at the $\Delta
\alpha /\alpha \sim O(10^{-5})$ level. Firstly, we are using $\sim 100$
lines, rather than $\sim 10,$ per system used in quasar studies. This gives
us a statistical advantage over the quasar studies. Secondly, the $q$ values
are much larger here since we are using more highly ionised species. Taken
together, this study should have more than an order-of-magnitude higher
sensitivity per system than the quasar studies, and we should be reaching
statistical accuracies below $10^{-6}$. Unfortunately, at present we are
limited by relatively poor laboratory wavelengths. In the future, this
limitation may be circumvented by comparing two white dwarfs
(or other stars) with different surface gravities. New measurements of the
\FeV\ and \NiV\ spectra could improve the limit by up to two orders of magnitude.

It is interesting to note that the gravitational redshift, $z=\Delta \phi
\approx 5\times 10^{-5}$, is the dominant contribution to the average total
redshift, $z_{abs}=7.78\times 10^{-5}$ for \FeV\ and $%
8.47\times 10^{-5}$ for \NiV. With improvement in
laboratory wavelengths, this system should also be able to provide a test of
the equivalence principle of general relativity in a `medium strength' field
with higher accuracy and
provide constraints on other variations of traditional `constants' driven by
scalar fields in the universe \cite{mbs,dent,BF}.

\onecolumngrid
\clearpage

\section{Supplemental Materials}

\begin{longtable*}{c@{\hskip 3mm}ccll@{\hskip 3mm}llc}
\caption{\label{tab:linesFe} Laboratory and astronomically observed lines in \FeV.
Laboratory wavelengths are taken from~\cite{ekberg75pscr} and have an estimated uncertainty of $\sim4$\,m\AA.
Observed wavelengths have a measurement uncertainty $\delta\lambda_\text{obs}$.
}\\
\hline\hline
$\lambda_\text{lab}$~(\AA) & $\lambda_\text{obs}$~(\AA) & $\delta\lambda_\text{obs}$~(m\AA) & \multicolumn{2}{c}{Lower Level} & \multicolumn{2}{c}{Upper Level} & $q$ (\cm) \\
\hline
\endfirsthead
\multicolumn{8}{c}
{\tablename\ \thetable: \FeV\ lines -- \textit{Continued from previous page}} \\
\hline\hline
$\lambda_\text{lab}$~(\AA) & $\lambda_\text{obs}$~(\AA) & $\delta\lambda_\text{obs}$~(m\AA) & \multicolumn{2}{c}{Lower Level} & \multicolumn{2}{c}{Upper Level} & $q$ (\cm) \\
\hline
\endhead
1234.648 & 1234.742 & 0.7 & $3d^3 ({^4}P) 4s$ & ${^3}P_2$ & $3d^3 ({^4}P) 4p$ & ${^3}S_1^o$ & 2484\\
1280.471 & 1280.569 & 0.5 & $3d^3 ({^2}G) 4s$ & ${^3}G_5$ & $3d^3 ({^2}D2) 4p$ & ${^3}F_4^o$ & 3238\\
1284.109 & 1284.207 & 1.6 & $3d^3 ({^4}P) 4s$ & ${^5}P_1$ & $3d^3 ({^4}P) 4p$ & ${^5}S_2^o$ & 3251\\
1285.918 & 1286.016 & 1.0 & $3d^3 ({^4}P) 4s$ & ${^3}P_1$ & $3d^3 ({^2}D2) 4p$ & ${^3}P_1^o$ & 3468\\
1288.169 & 1288.264 & 0.7 & $3d^3 ({^4}P) 4s$ & ${^5}P_2$ & $3d^3 ({^4}P) 4p$ & ${^5}S_2^o$ & 3054\\
1293.377 & 1293.481 & 0.6 & $3d^3 ({^2}G) 4s$ & ${^3}G_3$ & $3d^3 ({^2}D2) 4p$ & ${^3}F_2^o$ & 2634\\
1297.547 & 1297.645 & 0.6 & $3d^3 ({^4}P) 4s$ & ${^5}P_3$ & $3d^3 ({^4}P) 4p$ & ${^5}S_2^o$ & 2296\\
1300.608 & 1300.711 & 0.5 & $3d^3 ({^4}P) 4s$ & ${^5}P_3$ & $3d^3 ({^2}P) 4p$ & ${^1}D_2^o$ & 1391\\
1311.828 & 1311.932 & 0.5 & $3d^3 ({^2}P) 4s$ & ${^1}P_1$ & $3d^3 ({^2}D2) 4p$ & ${^1}D_2^o$ & 1797\\
1320.410 & 1320.514 & 0.3 & $3d^3 ({^2}H) 4s$ & ${^3}H_4$ & $3d^3 ({^2}H) 4p$ & ${^3}G_3^o$ & 3183\\
1321.341 & 1321.446 & 0.4 & $3d^3 ({^2}G) 4s$ & ${^3}G_4$ & $3d^3 ({^2}H) 4p$ & ${^3}H_5^o$ & 2508\\
1321.490 & 1321.594 & 0.3 & $3d^3 ({^2}G) 4s$ & ${^3}G_5$ & $3d^3 ({^2}H) 4p$ & ${^3}H_6^o$ & 2453\\
1321.850 & 1321.958 & 0.8 & $3d^3 ({^2}H) 4s$ & ${^3}H_4$ & $3d^3 ({^2}H) 4p$ & ${^3}G_4^o$ & 2819\\
1323.269 & 1323.373 & 0.3 & $3d^3 ({^2}H) 4s$ & ${^3}H_5$ & $3d^3 ({^2}H) 4p$ & ${^3}G_4^o$ & 2971\\
1325.781 & 1325.883 & 0.9 & $3d^3 ({^2}H) 4s$ & ${^3}H_5$ & $3d^3 ({^2}H) 4p$ & ${^3}G_5^o$ & 2874\\
1327.101 & 1327.207 & 1.5 & $3d^3 ({^2}D2) 4s$ & ${^1}D_2$ & $3d^3 ({^2}P) 4p$ & ${^1}P_1^o$ & 3137\\
1330.401 & 1330.508 & 0.2 & $3d^3 ({^2}H) 4s$ & ${^3}H_6$ & $3d^3 ({^2}H) 4p$ & ${^3}G_5^o$ & 2627\\
1331.185 & 1331.284 & 1.5 & $3d^3 ({^2}D2) 4s$ & ${^3}D_1$ & $3d^3 ({^2}D2) 4p$ & ${^3}P_0^o$ & 3514\\
1331.640 & 1331.744 & 0.5 & $3d^3 ({^2}D2) 4s$ & ${^1}D_2$ & $3d^3 ({^2}D2) 4p$ & ${^1}D_2^o$ & 2361\\
1350.535 & 1350.636 & 0.9 & $3d^3 ({^4}P) 4s$ & ${^3}P_1$ & $3d^3 ({^4}P) 4p$ & ${^3}D_2^o$ & 3430\\
1351.755 & 1351.856 & 1.5 & $3d^3 ({^4}F) 4s$ & ${^5}F_1$ & $3d^3 ({^4}F) 4p$ & ${^3}D_2^o$ & 3369\\
1354.847 & 1354.951 & 0.6 & $3d^3 ({^2}F) 4s$ & ${^1}F_3$ & $3d^3 ({^2}F) 4p$ & ${^1}F_3^o$ & 2656\\
1357.675 & 1357.777 & 0.7 & $3d^3 ({^2}F) 4s$ & ${^3}F_4$ & $3d^3 ({^2}F) 4p$ & ${^3}D_3^o$ & 2216\\
1358.567 & 1358.673 & 1.5 & $3d^3 ({^2}D2) 4s$ & ${^3}D_1$ & $3d^3 ({^2}D2) 4p$ & ${^3}D_2^o$ & 2933\\
1359.006 & 1359.113 & 0.8 & $3d^3 ({^2}D1) 4s$ & ${^3}D_3$ & $3d^3 ({^2}D1) 4p$ & ${^3}F_4^o$ & 3115\\
1361.447 & 1361.549 & 0.7 & $3d^3 ({^2}F) 4s$ & ${^1}F_3$ & $3d^3 ({^2}F) 4p$ & ${^1}G_4^o$ & 2630\\
1361.825 & 1361.929 & 0.5 & $3d^3 ({^2}F) 4s$ & ${^3}F_4$ & $3d^3 ({^2}F) 4p$ & ${^3}G_5^o$ & 3138\\
1362.864 & 1362.968 & 0.7 & $3d^3 ({^2}D2) 4s$ & ${^3}D_3$ & $3d^3 ({^2}D2) 4p$ & ${^3}D_3^o$ & 2948\\
1363.077 & 1363.181 & 0.5 & $3d^3 ({^4}F) 4s$ & ${^5}F_3$ & $3d^3 ({^4}F) 4p$ & ${^5}F_4^o$ & 3279\\
1363.642 & 1363.751 & 0.5 & $3d^3 ({^4}F) 4s$ & ${^5}F_4$ & $3d^3 ({^4}F) 4p$ & ${^5}F_5^o$ & 3066\\
1364.824 & 1364.923 & 0.8 & $3d^3 ({^4}F) 4s$ & ${^5}F_2$ & $3d^3 ({^4}F) 4p$ & ${^3}D_1^o$ & 1095\\
1364.984 & 1365.088 & 0.7 & $3d^3 ({^2}F) 4s$ & ${^3}F_2$ & $3d^3 ({^2}F) 4p$ & ${^3}D_3^o$ & 1900\\
1365.115 & 1365.222 & 0.7 & $3d^3 ({^4}F) 4s$ & ${^5}F_3$ & $3d^3 ({^4}F) 4p$ & ${^3}D_2^o$ & 2508\\
1365.571 & 1365.673 & 0.6 & $3d^3 ({^4}F) 4s$ & ${^5}F_2$ & $3d^3 ({^4}F) 4p$ & ${^5}F_3^o$ & 4776\\
1370.303 & 1370.411 & 1.8 & $3d^3 ({^2}H) 4s$ & ${^3}H_6$ & $3d^3 ({^2}H) 4p$ & ${^1}H_5^o$ & 2008\\
1370.947 & 1371.046 & 0.8 & $3d^3 ({^4}F) 4s$ & ${^5}F_1$ & $3d^3 ({^4}F) 4p$ & ${^5}F_2^o$ & 4189\\
1371.987 & 1371.046 & 0.8 & $3d^3 ({^2}D2) 4s$ & ${^3}D_1$ & $3d^3 ({^2}D2) 4p$ & ${^3}D_1^o$ & 1827\\
1373.587 & 1373.693 & 0.3 & $3d^3 ({^4}F) 4s$ & ${^5}F_4$ & $3d^3 ({^4}F) 4p$ & ${^5}F_4^o$ & 2611\\
1373.967 & 1374.072 & 0.5 & $3d^3 ({^4}P) 4s$ & ${^3}P_2$ & $3d^3 ({^4}P) 4p$ & ${^3}D_3^o$ & 1916\\
1374.116 & 1374.223 & 0.5 & $3d^3 ({^2}F) 4s$ & ${^3}F_3$ & $3d^3 ({^2}F) 4p$ & ${^3}G_4^o$ & 2554\\
1374.789 & 1374.897 & 1.0 & $3d^3 ({^2}D1) 4s$ & ${^3}D_2$ & $3d^3 ({^2}D1) 4p$ & ${^3}F_3^o$ & 2231\\
1376.337 & 1376.445 & 0.3 & $3d^3 ({^4}F) 4s$ & ${^5}F_5$ & $3d^3 ({^4}F) 4p$ & ${^5}F_5^o$ & 2275\\
1376.455 & 1376.560 & 0.5 & $3d^3 ({^4}F) 4s$ & ${^5}F_2$ & $3d^3 ({^4}F) 4p$ & ${^5}F_2^o$ & 3843\\
1378.560 & 1378.668 & 0.3 & $3d^3 ({^4}P) 4s$ & ${^5}P_3$ & $3d^3 ({^4}P) 4p$ & ${^5}D_4^o$ & 3246\\
1380.112 & 1380.219 & 0.6 & $3d^3 ({^4}F) 4s$ & ${^5}F_1$ & $3d^3 ({^4}F) 4p$ & ${^5}D_1^o$ &  3212\\
1384.055 & 1384.156 & 1.7 & $3d^3 ({^2}P) 4s$ & ${^3}P_1$ & $3d^3 ({^4}P) 4p$ & ${^3}D_2^o$ & 3245\\
1385.313 & 1385.421 & 0.9 & $3d^3 ({^4}F) 4s$ & ${^5}F_1$ & $3d^3 ({^4}F) 4p$ & ${^5}D_0^o$ & 3209\\
1387.938 & 1388.046 & 0.4 & $3d^3 ({^2}H) 4s$ & ${^3}H_6$ & $3d^3 ({^2}H) 4p$ & ${^3}I_7^o$ & 3186\\
1388.195 & 1388.296 & 1.5 & $3d^3 ({^4}P) 4s$ & ${^5}P_1$ & $3d^3 ({^4}P) 4p$ & ${^3}P_1^o$ & 3472\\
1388.328 & 1388.435 & 0.8 & $3d^3 ({^4}P) 4s$ & ${^5}P_1$ & $3d^3 ({^4}P) 4p$ & ${^5}D_2^o$ & 3157\\
1393.073 & 1393.181 & 1.0 & $3d^3 ({^4}P) 4s$ & ${^5}P_2$ & $3d^3 ({^4}P) 4p$ & ${^5}D_2^o$ & 2960\\
1394.272 & 1394.379 & 0.6 & $3d^3 ({^2}G) 4s$ & ${^3}G_4$ & $3d^3 ({^2}G) 4p$ & ${^3}F_3^o$ & 2685\\
1394.665 & 1394.778 & 0.7 & $3d^3 ({^2}G) 4s$ & ${^3}G_3$ & $3d^3 ({^2}G) 4p$ & ${^3}F_2^o$ & 2428\\
1397.106 & 1397.217 & 1.6 & $3d^3 ({^2}P) 4s$ & ${^3}P_1$ & $3d^3 ({^2}P) 4p$ & ${^3}S_1^o$ & 3650\\
1397.972 & 1398.079 & 0.9 & $3d^3 ({^4}P) 4s$ & ${^5}P_3$ & $3d^3 ({^4}P) 4p$ & ${^5}D_3^o$ & 2072\\
1400.243 & 1400.353 & 0.4 & $3d^3 ({^4}F) 4s$ & ${^3}F_2$ & $3d^3 ({^4}F) 4p$ & ${^3}F_2^o$ & 2782\\
1403.370 & 1403.479 & 0.8 & $3d^3 ({^2}G) 4s$ & ${^3}G_4$ & $3d^3 ({^2}G) 4p$ & ${^3}F_4^o$ & 1745\\
1404.260 & 1404.376 & 2.9 & $3d^3 ({^4}P) 4s$ & ${^3}P_0$ & $3d^3 ({^2}P) 4p$ & ${^3}D_1^o$ & 2018\\
1406.669 & 1406.781 & 0.5 & $3d^3 ({^4}F) 4s$ & ${^3}F_4$ & $3d^3 ({^4}F) 4p$ & ${^3}F_4^o$ & 2352\\
1407.248 & 1407.357 & 0.4 & $3d^3 ({^2}H) 4s$ & ${^1}H_5$ & $3d^3 ({^2}H) 4p$ & ${^1}I_6^o$ & 2587\\
1408.117 & 1408.230 & 0.7 & $3d^3 ({^4}F) 4s$ & ${^5}F_1$ & $3d^3 ({^4}F) 4p$ & ${^5}F_2^o$ & 3299\\
1409.026 & 1409.141 & 0.5 & $3d^3 ({^4}F) 4s$ & ${^5}F_3$ & $3d^3 ({^4}F) 4p$ & ${^5}D_2^o$ & 1339\\
1409.220 & 1409.334 & 0.4 & $3d^3 ({^4}F) 4s$ & ${^5}F_4$ & $3d^3 ({^4}F) 4p$ & ${^5}D_3^o$ & 1585\\
1416.219 & 1416.333 & 0.9 & $3d^3 ({^2}D2) 4s$ & ${^1}D_2$ & $3d^3 ({^2}D2) 4p$ & ${^1}F_3^o$ & 2285\\
1420.602 & 1420.713 & 0.6 & $3d^3 ({^2}G) 4s$ & ${^3}G_4$ & $3d^3 ({^2}G) 4p$ & ${^3}G_4^o$ & 2506\\
1422.481 & 1422.595 & 1.1 & $3d^3 ({^2}P) 4s$ & ${^3}P_1$ & $3d^3 ({^2}P) 4p$ & ${^3}D_2^o$ & 3381\\
1429.004 & 1429.118 & 0.7 & $3d^3 ({^2}G) 4s$ & ${^3}G_5$ & $3d^3 ({^2}G) 4p$ & ${^3}G_4^o$ & 2057\\
1430.309 & 1430.423 & 1.3 & $3d^3 ({^2}F) 4s$ & ${^1}F_3$ & $3d^3 ({^2}F) 4p$ & ${^1}D_2^o$ & 2573\\
1430.573 & 1430.688 & 0.4 & $3d^3 ({^4}F) 4s$ & ${^5}F_5$ & $3d^3 ({^4}F) 4p$ & ${^5}G_6^o$ & 3247\\
1430.751 & 1430.861 & 1.2 & $3d^3 ({^2}D2) 4s$ & ${^3}D_3$ & $3d^3 ({^4}P) 4p$ & ${^3}D_3^o$ & 1823\\
1440.528 & 1440.640 & 0.4 & $3d^3 ({^4}F) 4s$ & ${^5}F_4$ & $3d^3 ({^4}F) 4p$ & ${^5}G_5^o$ & 2704\\
1440.792 & 1440.904 & 1.0 & $3d^3 ({^4}P) 4s$ & ${^5}P_1$ & $3d^3 ({^4}P) 4p$ & ${^5}P_2^o$ & 1635\\
1441.049 & 1441.161 & 0.5 & $3d^3 ({^4}P) 4s$ & ${^5}P_3$ & $3d^3 ({^4}P) 4p$ & ${^5}P_3^o$ & 2565\\
1442.221 & 1442.335 & 0.5 & $3d^3 ({^2}G) 4s$ & ${^1}G_4$ & $3d^3 ({^2}G) 4p$ & ${^1}H_5^o$ & 2608\\
1446.618 & 1446.727 & 0.5 & $3d^3 ({^2}G) 4s$ & ${^3}G_5$ & $3d^3 ({^2}G) 4p$ & ${^3}H_6^o$ & 3148\\
1448.494 & 1448.604 & 0.6 & $3d^3 ({^2}G) 4s$ & ${^1}G_4$ & $3d^3 ({^2}G) 4p$ & ${^1}F_3^o$ & 3324\\
1448.846 & 1448.958 & 0.40 & $3d^3 ({^4}F) 4s$ & ${^5}F_3$ & $3d^3 ({^4}F) 4p$ & ${^5}G_4^o$ & 2277\\
1449.928 & 1450.037 & 0.7 & $3d^3 ({^2}F) 4s$ & ${^3}F_4$ & $3d^3 ({^2}F) 4p$ & ${^3}F_4^o$ & 2508\\
1453.618 & 1453.732 & 0.5 & $3d^3 ({^2}H) 4s$ & ${^1}H_5$ & $3d^3 ({^2}H) 4p$ & ${^1}H_5^o$ & 2063\\
1456.161 & 1456.278 & 0.5 & $3d^3 ({^4}F) 4s$ & ${^5}F_2$ & $3d^3 ({^4}F) 4p$ & ${^5}G_3^o$ & 1893\\
1456.285 & 1456.397 & 1.1 & $3d^3 ({^4}P) 4s$ & ${^5}P_2$ & $3d^3 ({^4}P) 4p$ & ${^5}P_1^o$ & 1828\\
1457.727 & 1457.840 & 1.0 & $3d^3 ({^4}P) 4s$ & ${^5}P_3$ & $3d^3 ({^4}P) 4p$ & ${^5}P_2^o$ & 1679\\
1460.726 & 1460.841 & 0.8 & $3d^3 ({^4}F) 4s$ & ${^5}F_4$ & $3d^3 ({^4}F) 4p$ & ${^5}G_4^o$ & 1609\\
1462.631 & 1462.747 & 0.5 & $3d^3 ({^4}F) 4s$ & ${^5}F_1$ & $3d^3 ({^4}F) 4p$ & ${^5}G_2^o$ & 1534\\
1464.876 & 1464.990 & 1.1 & $3d^3 ({^2}F) 4s$ & ${^3}F_2$ & $3d^3 ({^2}F) 4p$ & ${^3}F_2^o$ & 1810\\
1466.649 & 1466.763 & 0.6 & $3d^3 ({^2}G) 4s$ & ${^3}G_4$ & $3d^3 ({^2}G) 4p$ & ${^3}H_5^o$ & 1935\\
1468.911 & 1469.022 & 0.8 & $3d^3 ({^4}F) 4s$ & ${^5}F_2$ & $3d^3 ({^4}F) 4p$ & ${^5}G_2^o$ & 1188\\
1469.000 & 1469.115 & 0.4 & $3d^3 ({^2}H) 4s$ & ${^3}H_6$ & $3d^3 ({^2}H) 4p$ & ${^3}H_6^o$ & 2401\\
1472.098 & 1472.214 & 0.5 & $3d^3 ({^2}H) 4s$ & ${^3}H_5$ & $3d^3 ({^2}H) 4p$ & ${^3}H_5^o$ & 2252\\
1472.512 & 1472.627 & 0.6 & $3d^3 ({^2}H) 4s$ & ${^3}H_4$ & $3d^3 ({^2}H) 4p$ & ${^3}H_4^o$ & 2372\\
1475.604 & 1475.715 & 1.1 & $3d^3 ({^2}G) 4s$ & ${^3}G_5$ & $3d^3 ({^2}G) 4p$ & ${^3}H_5^o$ & 1485\\
1479.471 & 1479.589 & 0.5 & $3d^3 ({^2}G) 4s$ & ${^3}G_4$ & $3d^3 ({^2}G) 4p$ & ${^3}H_4^o$ & 1132\\
1496.266 & 1496.384 & 0.7 & $3d^3 ({^2}G) 4s$ & ${^1}G_4$ & $3d^3 ({^2}G) 4p$ & ${^3}F_4^o$ & 1732\\
1530.439 & 1530.572 & 1.8 & $3d^3 ({^2}D2) 4s$ & ${^1}D_2$ & $3d^3 ({^2}D2) 4p$ & ${^1}P_1^o$ & 1967\\
1543.234 & 1543.362 & 1.0 & $3d^3 ({^4}F) 4s$ & ${^3}F_2$ & $3d^3 ({^4}F) 4p$ & ${^3}D_1^o$ & 901\\
1554.219 & 1554.345 & 0.6 & $3d^3 ({^4}F) 4s$ & ${^3}F_4$ & $3d^3 ({^4}F) 4p$ & ${^3}D_3^o$ & 1234\\
\hline
\end{longtable*}

\clearpage

\begin{longtable*}{c@{\hskip 3mm}ccll @{\hskip 3mm} llc}
\caption{\label{tab:linesNi} Laboratory and astronomically observed lines in \NiV.
Laboratory wavelengths are taken from~\cite{raassen77physc} and we estimate their uncertainty at $\sim7$\,m\AA.
Observed wavelengths have a measurement uncertainty $\delta\lambda_\text{obs}$.
Where $q$ values have been omitted, they could not be unambiguously identified in the \emph{ab initio} calculations. Where lines were blended in the laboratory data, they are marked with an asterisk in the $\lambda_\text{lab}$ column.}\\
\hline\hline
$\lambda_\text{lab}$~(\AA) & $\lambda_\text{obs}$~(\AA) & $\delta\lambda_\text{obs}$~(m\AA) & \multicolumn{2}{c}{Lower Level} & \multicolumn{2}{c}{Upper Level} & $q$ (\cm) \\
\hline
\endfirsthead
\multicolumn{8}{c}
{\tablename\ \thetable: \NiV\ lines -- \textit{Continued from previous page}} \\
\hline\hline
$\lambda_\text{lab}$~(\AA) & $\lambda_\text{obs}$~(\AA) & $\delta\lambda_\text{obs}$~(m\AA) & \multicolumn{2}{c}{Lower Level} & \multicolumn{2}{c}{Upper Level} & $q$ (\cm) \\
\hline
\endhead
1202.423 & 1202.515 & 1.0 & $3d^5 ({^4}G) 4s$ & ${^5}G_4$ & $3d^5 ({^4}G) 4p$ & ${^3}F_3^o$ & 3378\\
1227.480 & 1227.635 & 0.3 & $3d^5 ({^2}F1) 4s$ & ${^3}F_4$& $3d^5 ({^2}H) 4p$ & ${^3}G_4^o$& \\
1232.801 & 1232.905 & 0.7 & $3d^5 ({^2}H) 4s$ & ${^1}H_5$ & $3d^5 ({^2}H) 4p$ & ${^1}H_5^o$ & \\
1235.831 & 1235.936 & 0.5 & $3d^5 ({^2}I) 4s$ & ${^1}I_6$ & $3d^5 ({^4}F) 4p$ & ${^5}G_6^o$ & \\
1242.072 & 1242.180 & 0.6 & $3d^5 ({^2}I) 4s$ & ${^1}I_6$ & $3d^5 ({^2}I) 4p$ & ${^3}H_5^o$ & 3964\\
1244.174 & 1244.286 & 0.2 & $3d^5 ({^6}S) 4s$ & ${^7}S_3$ & $3d^5 ({^6}S) 4p$ & ${^7}P_4^o$ & 4234\\
1249.520 & 1249.631 & 0.5 & $3d^5 ({^4}G) 4s$ & ${^5}G_5$ & $3d^5 ({^4}G) 4p$ & ${^5}F_4^o$ & 2901\\
1250.384 & 1250.498 & 0.2 & $3d^5 ({^4}G) 4s$ & ${^5}G_6$ & $3d^5 ({^4}G) 4p$ & ${^5}H_7^o$ & 4152\\
1251.821 & 1251.933 & 0.4 & $3d^5 ({^4}G) 4s$ & ${^3}G_5$ & $3d^5 ({^4}G) 4p$ & ${^3}G_5^o$ & 3324\\
1252.155* & 1252.269 & 0.4 & $3d^5 ({^4}G) 4s$ & ${^5}G_6$ & $3d^5 ({^4}F) 4p$ & ${^5}F_5^o$ & \\
1252.267 & 1252.374 & 1.3 & $3d^5 ({^2}I) 4s$ & ${^3}I_6$ & $3d^5 ({^2}I) 4p$ & ${^3}H_6^o$ & 3343\\
1254.187 & 1254.300 & 1.3 & $3d^5 ({^2}F1) 4s$ & ${^1}F_3$ & $3d^5 ({^2}F1) 4p$ & ${^1}D_2^o$ & \\
1257.620 & 1257.735 & 0.2 & $3d^5 ({^4}G) 4s$ & ${^5}G_5$ & $3d^5 ({^4}G) 4p$ & ${^5}H_6^o$ & 3872\\
1261.330 & 1261.445 & 0.5 & $3d^5 ({^2}F1) 4s$ & ${^3}F_4$ & $3d^5 ({^2}F1) 4p$ & ${^3}G_5^o$ & \\
1261.745 & 1261.859 & 0.3 & $3d^5 ({^4}D) 4s$ & ${^5}D_4$ & $3d^5 ({^4}D) 4p$ & ${^5}F_5^o$ & 4269\\
1264.518 & 1264.635 & 0.2 & $3d^5 ({^4}G) 4s$ & ${^5}G_6$ & $3d^5 ({^4}G) 4p$ & ${^5}H_5^o$ & 3059\\
1266.395 & 1266.507 & 0.3 & $3d^5 ({^4}G) 4s$ & ${^5}G_4$ & $3d^5 ({^4}G) 4p$ & ${^5}H_5^o$ & 3088\\
1266.859 & 1266.969 & 0.7 & $3d^5 ({^4}P) 4s$ & ${^5}P_1$ & $3d^5 ({^4}D) 4p$ & ${^5}P_2^o$ & \\
1267.275 & 1267.401 & 0.6 & $3d^5 ({^2}I) 4s$ & ${^3}I_6$ & $3d^5 ({^2}I) 4p$ & ${^3}I_7^o$ & 4423\\
1267.803 & 1267.910 & 0.6 & $3d^5 ({^2}I) 4s$ & ${^3}I_7$ & $3d^5 ({^2}I) 4p$ & ${^3}I_7^o$ & 4182\\
1270.677 & 1270.791 & 0.3 & $3d^5 ({^2}I) 4s$ & ${^3}I_7$ & $3d^5 ({^2}I) 4p$ & ${^3}K_8^o$ & 4076\\
1273.198 & 1273.311 & 0.3 & $3d^5 ({^4}G) 4s$ & ${^5}G_3$ & $3d^5 ({^4}G) 4p$ & ${^5}H_4^o$ & 2715\\
1276.415 & 1276.533 & 0.5 & $3d^5 ({^4}D) 4s$ & ${^5}D_3$ & $3d^5 ({^4}D) 4p$ & ${^5}F_4^o$ & 3073\\
1276.945 & 1277.052 & 0.2 & $3d^5 ({^6}S) 4s$ & ${^7}S_3$ & $3d^5 ({^6}S) 4p$ & ${^7}P_2^o$ & 1942\\
1279.708 & 1279.818 & 0.4 & $3d^5 ({^4}G) 4s$ & ${^5}G_2$ & $3d^5 ({^4}G) 4p$ & ${^5}H_3^o$ & 2225\\
1282.247* & 1282.309 & 0.9 & $3d^5 ({^2}I) 4s$ & ${^3}I_6$ & $3d^5 ({^2}I) 4p$ & ${^3}I_6^o$ & 2909\\
1282.247* & 1282.383 & 0.8 & $3d^5 ({^2}I) 4s$ & ${^3}I_5$ & $3d^5 ({^2}I) 4p$ & ${^3}I_6^o$ & 3023\\
1287.576 & 1287.671 & 1.1 & $3d^5 ({^2}D3) 4s$ & ${^3}D_3$ & $3d^5 ({^2}D3) 4p$ & ${^3}F_4^o$ & \\
1295.286 & 1295.391 & 0.9 & $3d^5 ({^4}P) 4s$ & ${^5}P_3$ & $3d^5 ({^4}G) 4p$ & ${^5}F_3^o$ & 3561\\
1298.733 & 1298.829 & 1.0 & $3d^5 ({^2}G2) 4s$ & ${^3}G_5$ & $3d^5 ({^2}H) 4p$ & ${^3}I_6^o$ & \\
1300.981 & 1301.086 & 0.4 & $3d^5 ({^6}S) 4s$ & ${^5}S_2$ & $3d^5 ({^6}S) 4p$ & ${^5}P_1^o$ & 3651\\
1302.380 & 1302.490 & 1.1 & $3d^5 ({^4}P) 4s$ & ${^5}P_3$ & $3d^5 ({^4}P) 4p$ & ${^5}S_2^o$ & 3062\\
1307.595 & 1307.702 & 0.3 & $3d^5 ({^6}S) 4s$ & ${^5}S_2$ & $3d^5 ({^6}S) 4p$ & ${^5}P_2^o$ & 3304\\
1310.249 & 1310.354 & 1.2 & $3d^5 ({^4}F) 4s$ & ${^3}F_4$ & $3d^5 ({^4}F) 4p$ & ${^3}G_5^o$ & \\
1311.106 & 1311.215 & 0.4 & $3d^5 ({^4}G) 4s$ & ${^5}G_5$ & $3d^5 ({^4}G) 4p$ & ${^5}G_5^o$ & 2343\\
1313.303 & 1313.411 & 0.4 & $3d^5 ({^4}G) 4s$ & ${^5}G_4$ & $3d^5 ({^4}G) 4p$ & ${^5}G_4^o$ & 2256\\
1314.330 & 1314.437 & 0.4 & $3d^5 ({^4}G) 4s$ & ${^5}G_3$ & $3d^5 ({^4}G) 4p$ & ${^5}G_3^o$ & 2260\\
1317.436 & 1317.547 & 0.3 & $3d^5 ({^2}I) 4s$ & ${^1}I_6$ & $3d^5 ({^2}I) 4p$ & ${^1}K_7^o$ & 3483\\
1317.962 & 1318.080 & 0.9 & $3d^5 ({^2}F1) 4s$ & ${^1}F_3$ & $3d^5 ({^2}F1) 4p$ & ${^3}G_4^o$ & \\
1318.148 & 1318.251 & 0.8 & $3d^5 ({^4}P) 4s$ & ${^5}P_3$ & $3d^5 ({^4}G) 4p$ & ${^5}F_3^o$ & 3561\\
1318.513 & 1318.614 & 0.3 & $3d^5 ({^6}S) 4s$ & ${^5}S_2$ & $3d^5 ({^6}S) 4p$ & ${^5}P_3^o$ & 2721\\
1329.372 & 1329.476 & 0.4 & $3d^5 ({^4}G) 4s$ & ${^3}G_4$ & $3d^5 ({^4}G) 4p$ & ${^3}H_5^o$ & 3306\\
1336.157 & 1336.264 & 0.6 & $3d^5 ({^4}G) 4s$ & ${^3}G_5$ & $3d^5 ({^4}G) 4p$ & ${^3}H_6^o$ & 3042\\
1342.177 & 1342.288 & 0.7 & $3d^5 ({^4}G) 4s$ & ${^3}G_5$ & $3d^5 ({^4}G) 4p$ & ${^3}F_4^o$ & 3450\\
\hline
\end{longtable*}


\begin{thebibliography}{99}
\bibitem{mbs} J. Magueijo, J. D. Barrow and H. Sandvik, 
Phys. Lett. B \textbf{549}, 284 (2002).

\bibitem{flambaum08aipconf} V. V. Flambaum and E. V. Shuryak, AIP Conf.
Proc. \textbf{995}, 1 (2008), arXiv:physics/0701220.

\bibitem{bek} J. D. Bekenstein, Phys. Rev. D 25, 1527 (1982).

\bibitem{bsbm} H. B. Sandvik, J. D. Barrow and J. Magueijo, Phys. Rev. Lett. 
\textbf{88}, 031302 (2002); J. D. Barrow, H. B. Sandvik and J. Magueijo, Phys.
Rev. D \textbf{65}, 063504 (2002).

\bibitem{blip} J. D. Barrow and S. Z. W. Lip, Phys. Rev. D \textbf{85}, 023514
(2012).

\bibitem{bauch02prd} A. Bauch and S. Weyers, Phys. Rev. D \textbf{65},
081101 (2002).

\bibitem{ferrell07pra} S. J. Ferrell, A. Cing{\"{o}}z, A. Lapierre, A.-T.
Nguyen, N. Leefer, D. Budker, V. V. Flambaum, S. K. Lamoreaux, and J. R. Torgerson,
Phys. Rev. A \textbf{76}, 062104 (2007).

\bibitem{fortier07prl} T. M. Fortier, N. Ashby, J. C. Bergquist, M. J.
Delaney, S. A. Diddams, T. P. Heavner, L. Hollberg, W. M. Itano, S. R.
Jefferts, K. Kim, F. Levi, L. Lorini, W. H. Oskay, T. E. Parker, J. Shirley,
and J. E. Stalnaker, Phys. Rev. Lett. \textbf{98}, 070801 (2007).

\bibitem{blatt08prl} S. Blatt, A. D. Ludlow, G. K. Campbell, J. W. Thomsen,
T. Zelevinsky, M. M. Boyd, J. Ye, X. Baillard, M. Fouche, R. Le~Targat, A.
Brusch, P. Lemonde, M. Takamoto, F.-L. Hong, H. Katori, and V. V. Flambaum,
Phys. Rev. Lett. \textbf{100}, 140801 (2008).

\bibitem{BShaw} J.D. Barrow and D.J. Shaw, Phys. Rev. D \textbf{78}, 067304
(2008).

\bibitem{guena12prl} J. Gu{\'{e}}na, M. Abgrall, D. Rovera, P. Rosenbusch,
M. E. Tobar, P. Laurent, A. Clarion, and S. Bize, Phys. Rev. Lett. \textbf{%
109}, 080801 (2012).

\bibitem{leefer13arxiv}
N. Leefer, C. T. M. Weber, A. Cing{\"o}z, J. R. Torgerson, and D. Budker,
``New limits on variation of the fine-structure constant using atomic dysprosium'',
arXiv:1304.6940 (2013).

\bibitem{reid88apj}
N. Reid and G. Wegner, Astrophys. J. \textbf{335}, 953 (1988).

\bibitem{preval13mnras} S. P. Preval, M. A. Barstow, J. B. Holberg, and N.
J. Dickinson, Mon. Not. R. Astron. Soc., submitted (2013).

\bibitem{chayer95apj} P. Chayer, S. Vennes, A. K. Pradhan, P. Thejll, A.
Beauchamp, G. Fontaine, and F. Wesemael, Astrophys. J. \textbf{454}, 429
(1995).


\bibitem{dzuba96pra} V. A. Dzuba, V. V. Flambaum, and M. G. Kozlov, Phys.
Rev. A \textbf{54}, 3948 (1996).

\bibitem{berengut05pra} J. C. Berengut, V. V. Flambaum, and M. G. Kozlov,
Phys. Rev. A \textbf{72}, 044501 (2005).

\bibitem{berengut06pra} J. C. Berengut, V. V. Flambaum, and M. G. Kozlov,
Phys. Rev. A \textbf{73}, 012504 (2006).

\bibitem{berengut08jpb} J. C. Berengut, V. V. Flambaum, and M. G. Kozlov, J.
Phys. B \textbf{41}, 235702 (2008).

\bibitem{sugar85jpcrd} J. Sugar and C. Corliss, J. Phys. Chem. Ref. Data 
\textbf{14}, Suppl. 2, 1 (1985).

\bibitem{berengut11pra1} J. C. Berengut, Phys. Rev. A \textbf{84}, 052520
(2011).


\bibitem{holberg03aspcs} J. B. Holberg, M. A. Barstow, I. Hubeny, M. S.
Sahu, F. C. Bruhweiler, and W. B. Landsman, ASP Conf. Ser. \textbf{291}, 383
(2003).

\bibitem{kurucz11cjp} R. L. Kurucz, Can. J. Phys. \textbf{89}, 417 (2011),
http://kurucz.harvard.edu.

\bibitem{kentuckylinelist} P. van Hoof, ``Atomic line list V2.05,''
http://www.pa.uky.edu/{$\sim$}peter/newpage.

\bibitem{ekberg75pscr} J. O. Ekberg, Phys. Scr. \textbf{12}, 42 (1975).

\bibitem{raassen77physc} A. J. J. Raassen and Th. A. M. van Kleef, Physica C 
\textbf{85}, 180 (1977).

\bibitem{raassen76physc} A. J. J. Raassen, Th. A. M. van Kleef, and B. C.
Metsch, Physica C \textbf{84}, 133 (1976).

\bibitem{dzuba99prl} V. A. Dzuba, V. V. Flambaum, and J. K. Webb, Phys. Rev.
Lett. \textbf{82}, 888 (1999).

\bibitem{webb99prl} J. K. Webb, V. V. Flambaum, C. W. Churchill, M. J.
Drinkwater, and J. D. Barrow, Phys. Rev. Lett. \textbf{82}, 884 (1999).

\bibitem{dent} T. Dent, Phys. Rev. Lett. \textbf{101}, 041102 (2008).

\bibitem{BF} J. C. Berengut and V. V. Flambaum, Europhys. Lett. \textbf{97},
20006 (2012)
\end{thebibliography}
\end{document}